# An RNA condensate model for the origin of life.

Jacob L. Fine[1,2] & Alan M. Moses[3,4,5]

[1]Donnelly Centre, University of Toronto, Toronto
[2]Molecular Genetics, University of Toronto, Toronto
[3]Cell & Systems Biology, University of Toronto, Toronto
[4]Computer Science, University of Toronto, Toronto
[5]Ecology and Evolutionary Biology, University of Toronto, Toronto

### *Graphical Abstract*

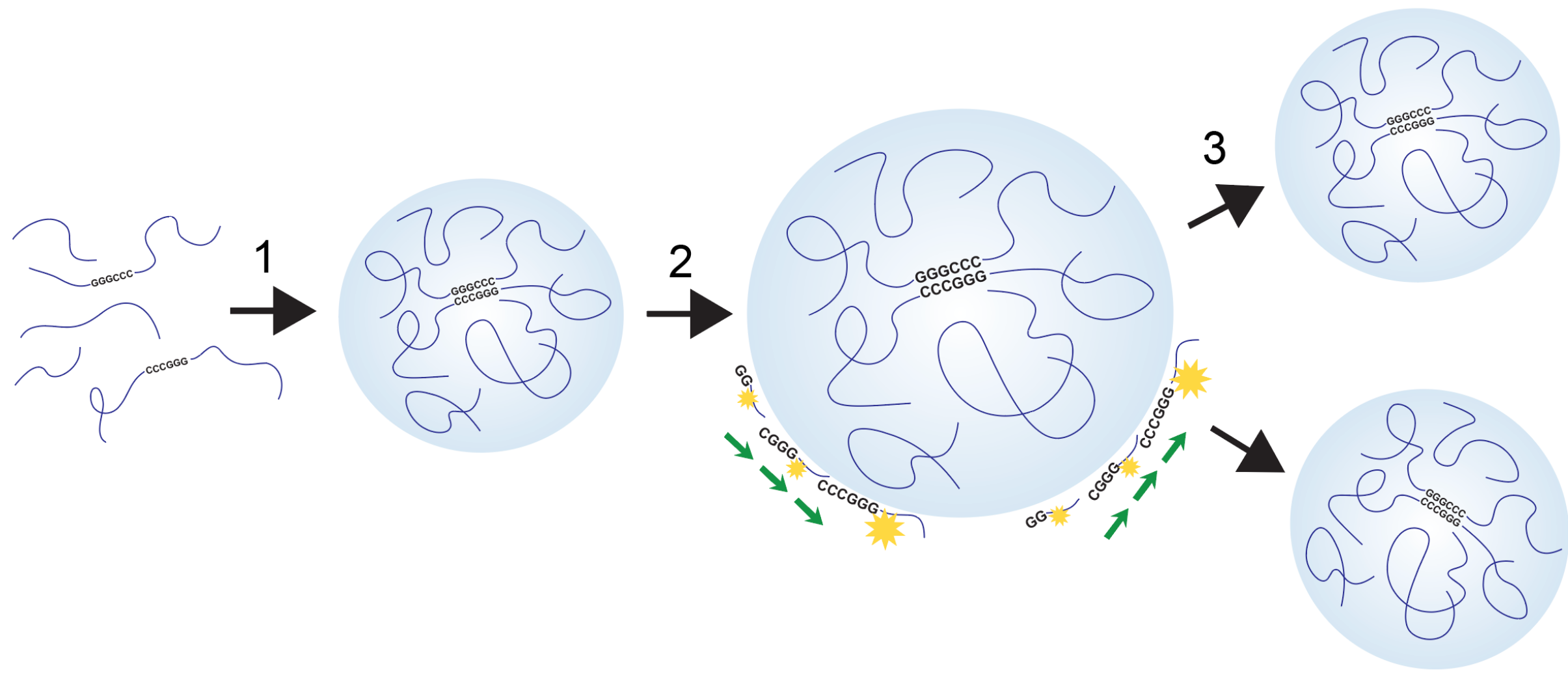

Proposed mechanism of compartmentalization and self-replication via an RNA condensate: (1) condensation, (2) RNA-templated polymerization (green arrows) and growth, and (3) division.

### *Research Highlights*

1. RNA condensates may have served as self-replicating catalysts for the origin of life and the RNA World.
2. Catalytic RNA condensates formed by short and low-complexity RNAs lower the error threshold and unify replication and compartmentalization.
3. RNA condensates capable of self-templated polymerization may spontaneously begin a ‘condensate chain reaction’ leading to natural selection
4. Within a standard polymer physics framework, key biophysical parameters are defined for condensate-dependent self-templated RNA polymerization, leading to a thermodynamic and information theoretic framework for the origin of life.
5. These hypotheses about the origin of life can readily be tested by experiment or simulations.

### *Abstract*

The RNA World hypothesis predicts that self-replicating RNAs evolved before DNA genomes and coded proteins. Despite widespread support for the RNA World, self-replicating RNAs have yet to be identified in a natural context, leaving a key ‘missing link’ for this explanation of the origin of life. Inspired by recent work showing that condensates of charged polymers are capable of catalyzing chemical reactions, we consider a catalytic RNA condensate as a candidate for the self-replicating RNA. Specifically, we propose that short, low-complexity RNA polymers formed catalytic condensates capable of templated RNA polymerization. Because the condensate properties depend on the RNA sequences, RNAs that formed condensates with improved polymerization and demixing capacity would be amplified, leading to a 'condensate chain reaction' and evolution by natural selection. Many of the needed properties of this self-replicating RNA condensate have been realized experimentally in recent studies and our predictions could be tested with current experimental and theoretical tools. Our theory addresses central problems in the origins of life: (i) the origin of compartmentalization, (ii) the error threshold for the accuracy of templated replication, (iii) the free energy cost of maintaining an information-rich population of replicating RNA polymers. Furthermore, we note that the extant nucleolus appears to satisfy many of the requirements of an evolutionary relic for the model we propose. More generally, we suggest that future work on the origin of life would benefit from condensate-centric biophysical models of RNA evolution.

## *Introduction*

The joint informatic and catalytic properties of RNA suggest that life originated in an RNA-dominated context with self-replicating RNAs. This view, known as the RNA World hypothesis[1] has gained widespread support since the discovery of catalytic RNA,[2,3] and the finding that peptide synthesis is catalyzed by a ribozyme.[4] From the latter it follows that catalytic RNA must predate the emergence of the genetic code and coded proteins altogether[5–7] and other evidence also supports the view that RNA arose before DNA genomes and coded proteins.[2–4,8,9]

Despite the support for the RNA World, the exact mechanism by which the first RNAs arose in a 'pre-RNA World' remains unclear.[5] For a self-replicating RNA to evolve, a barrier between the individual and its environment is required, along with a mechanism that accounts for the synthesis of new RNAs from parent generations.[5] In other words, the origins of the RNA World is a function of (i) the nature of the original catalyst of prebiotic RNA self-replication and (ii) the original mechanism of RNA compartmentalization to form the division between the system and its surroundings.

To address self-replication, it has been suggested that an RNA-dependent RNA polymerase (RdRp) composed of RNA arose before the protein replicases of present life that drove early RNA replication.[10] While success has been attained in synthesizing artificial RdRp ribozymes[11–15] to our knowledge there are no known naturally occurring enzymes of this character and artificial RdRp ribozymes have relatively long (typically at least ~100-150 bases) and stable structures that are highly sensitive to the sequences of the molecules that encode them. This makes it necessary to explain how RNA underwent self-replication and selection in the astronomically large sequence space, before the origins of a functional RdRp ribozyme of sufficient length, fidelity and catalytic efficiency.

This notion was formalized by the error threshold,[16] where longer polymers require lower mutation rates to replicate with sufficient copying accuracy to avoid excessive information loss. To understand this relationship, consider an RNA 'master sequence' of length $N$ undergoing self-replication with a mutation rate $r$, where mutations occur independently at each base. The 'fitness' of the master sequence, $\sigma$, is defined as the number of identical copies made per mutant copy. If the error rate $r$ exceeds $\ln(\sigma)/N$ (Equation 1) then an 'error catastrophe' occurs,[16] where the abundance of the master sequence rapidly decreases relative to the growing population of mutant sequences.

$$r < \frac{\ln(\sigma)}{N}$$

(Equation 1)

Thus, the error rate of RNA self-replication in the RNA world is bounded by Equation 1. In practice, for a ribozyme requiring the exact specification of $N$ nucleotides, the error threshold can be taken[17] as $1/N$ although higher thresholds may be permissible.[18] For ribozymes of ~100 bases, this seems difficult to obtain ($r < 1\%$) without a relatively long, highly ordered RdRp ribozyme in the first place. This problem is known as Eigen's paradox and is considered a major challenge for the spontaneous origin of life.

In addition to replicative catalysis, it is widely held that compartmentalization of prebiotic chemistry was critical. Because in all extant cells, compartmentalization depends on membranes, current work emphasizes the synthesis of protocells composed of simple lipids to model prebiotic events, particularly, the replication and evolution of membrane-encapsulated RNAs.[19–21] While important progress has been made, for instance, the production of RNA polymers up to ~200 bases within protocells,[20] template-directed RNA self-replication coupled with dividing protocells over multiple generations has yet to be achieved.[5]

A possible simpler mode of 'pre-membrane' compartmentalization is condensation. Condensation is a spontaneous process in which polymers in solution separate into a dilute phase and a dense phase, constituting demixing.[22] This gives rise to cellular compartments known as 'membrane-less organelles' which are often rich in RNAs and low-complexity proteins.[23] Indeed, condensates have been considered in the context of the origin of life[24] and coacervates (condensates formed via phase separation) were long considered a possible physical substrate for the origin of life as initially proposed by Oparin.[25] In the context of current cellular life, it has been suggested that condensates function to improve the kinetics of biochemical reactions, for instance, by increasing the frequency of molecular collisions that occur in the correct orientation.[26] This has made condensates a subject of recent interest in origin of life theory, as they can serve to localize chemical reactions and sequester molecules for faster reaction kinetics.[27]

Condensation is thermodynamically favorable when polymer-polymer interactions are sufficiently large such that they offset solute-solute and solute-polymer interactions. In the simplest physical model of this process with a binary mixture of polymer and solvent, this is captured by the magnitude of the interaction parameter denoted $\chi$, expressed in terms of the energies of solvent-polymer ($sp$), polymer-polymer ($pp$), and solvent-solvent ($ss$) interactions (Equation 2).[28] If the quantity $\chi$ exceeds a critical value, mixing is thermodynamically unfavorable, so demixing occurs. This free energy difference is known as the free energy of mixing ($\Delta G_{mix}$) and is discussed later.

$$\chi = \frac{2w_{sp} - (w_{pp} + w_{ss})}{k_B T}$$

(Equation 2)

The free energy stored in the condensate is in principle available as a potential to do work. Indeed, recent discoveries in polymer and condensate biophysics indicate that condensates of charged polymers can function as catalysts of redox[29] and hydrolysis reactions, including of ATP,[30] and can generate electric potentials and electrochemical gradients[31–33] capable of doing work. Thus, remarkably, membrane-less bodies can exhibit physical membrane-like properties beyond simple compartmentalization. Since the individual simple polymers in a condensate lack catalytic activity themselves, catalysis is an emergent property of some populations of polymers that undergo condensation.

Intriguingly, a spontaneously forming condensate capable of template-based hydrolysis of RNA polymerization provides (in principle) a mechanism for both compartmentalization and self-replication. In other words, we contemplate the possibility that the condensate itself was a self-replicative catalyst during the origin of life, rather than a prebiotic compartment for ribozymes and other reactions to evolve by individual molecular catalysts. Here, we explore the implications of this idea further based on recent experimental studies and a simple polymer physics model of condensation. We are surprised to discover that with only these simple assumptions, other key challenges for the origin of life are addressed, and that a condensate chain reaction could apparently lead to evolution by natural selection.

***Spontaneous evolution by natural selection through a 'Condensate Chain Reaction'***

We assume that in the pre-RNA World, non-templated abiotic modes of polymerization produced populations of RNA polymers prior to the origins of templated replication.[5] We propose (Figure 1) that these RNAs underwent phase separation once $\chi$ exceeded its critical value (Figure 1, black arrow 1), which depends on the RNA sequence. Polymerization (Figure 1, green arrows) occurs at the surface of the dense phase, leading to increased polymer concentration, which favors increased condensate growth. We imagine that catalysis is facilitated by electric fields created by the condensate (discussed further below) that orient the RNA to facilitate nucleophilic attack, as in extant polymerases. For condensate division (Figure 1, black arrow 3), we hypothesize that physical forces may have acted to shear apart condensates after sufficient growth, resembling cycles of model protocell growth and division.[19] Since $\chi$ is also a function of temperature ($\chi \propto 1/T$), we also consider that temperature cycles in the prebiotic earth governed the dynamics of condensate growth and division, just as temperature cycles control PCR.

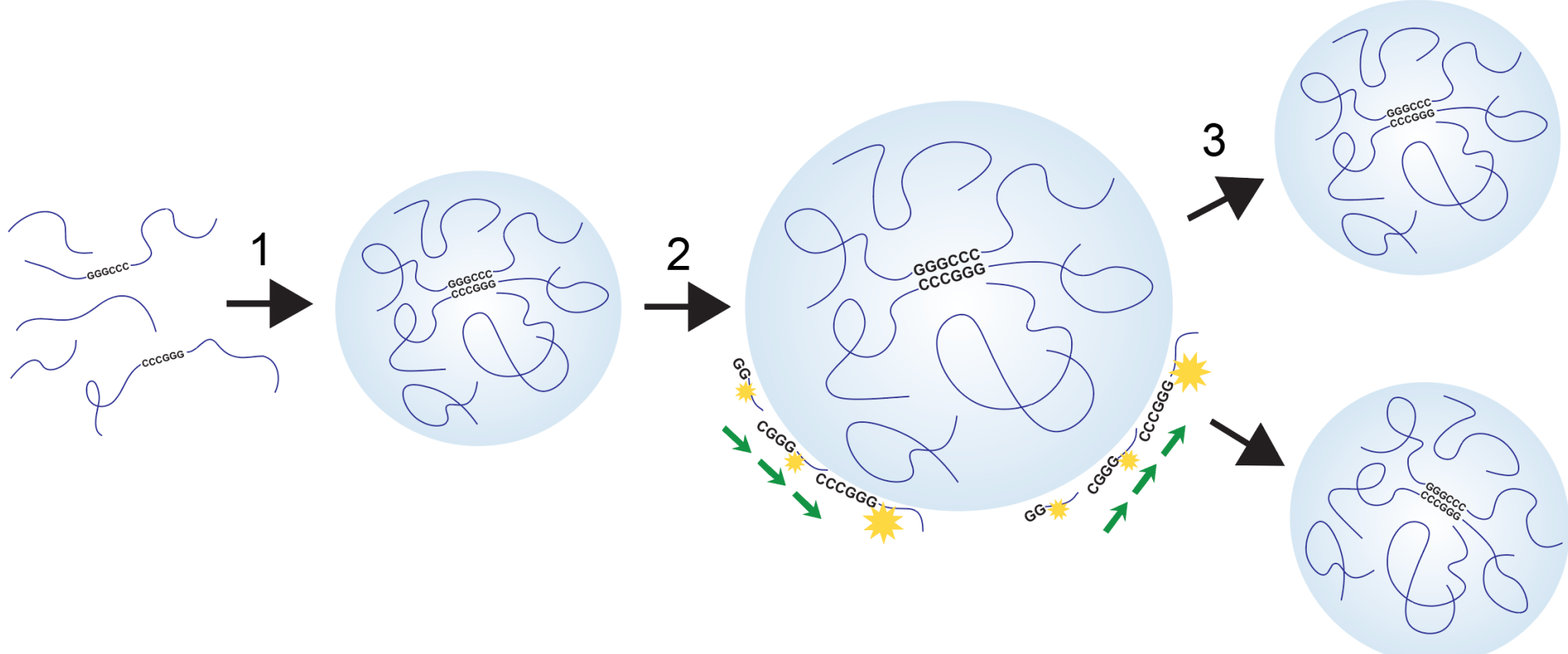

**Figure 1.** Proposed mechanism of compartmentalization and self-replication via an RNA condensate: (1) Polymers form compartments through spontaneous condensation. (2) The RNA condensate enables self-templated polymerization (green arrows) along the surface between the dilute (solvent rich) and dense (polymer rich) droplets. These new polymers are incorporated into the condensate, which leads to growth. (3) The large RNA condensate breaks up into smaller droplets either through physical shearing or due to changes in temperature.

Once the condensate breaks apart, the ‘offspring’ condensates can undergo further growth and division, resulting in a positive feedback loop that produces a ‘condensate chain reaction’ (Figure 2). Because the biophysical properties governing condensation, self-templated polymerization and division depend on the RNA sequences, errors in polymerization will lead to RNAs with slight differences in these properties. In other words, mutation will create variation in condensate phenotypes. For example, if the new RNA sequences are sufficiently mutated (red stars in Figure 2) so that the new polymers no longer encode $\chi$ beyond its critical value, de-mixing would no longer occur, and those RNAs would no longer form compartments.

The ‘condensate chain reaction’ with errors in replication appears sufficient to yield natural selection: since the self-replication process is a chain-reaction, we expect exponential amplification of RNAs that continue to de-mix and self-template, and extinction of lineages of condensates harboring mutations that reduce the capacity to do so. This natural selection is predicted to act primarily on (i) the capacity to elicit phase separation mediated by $\chi$, (ii) the capacity of condensates to divide and (iii) the polymerization ability of the condensates to copy RNA polymers that favor condensate formation. Notably, if base composition is sufficient for de-mixing capacity, then the condensate chain reaction may even begin without a requirement to correctly order the bases in the RNA. A formal framework for the impact of base composition on phase separation is discussed later.

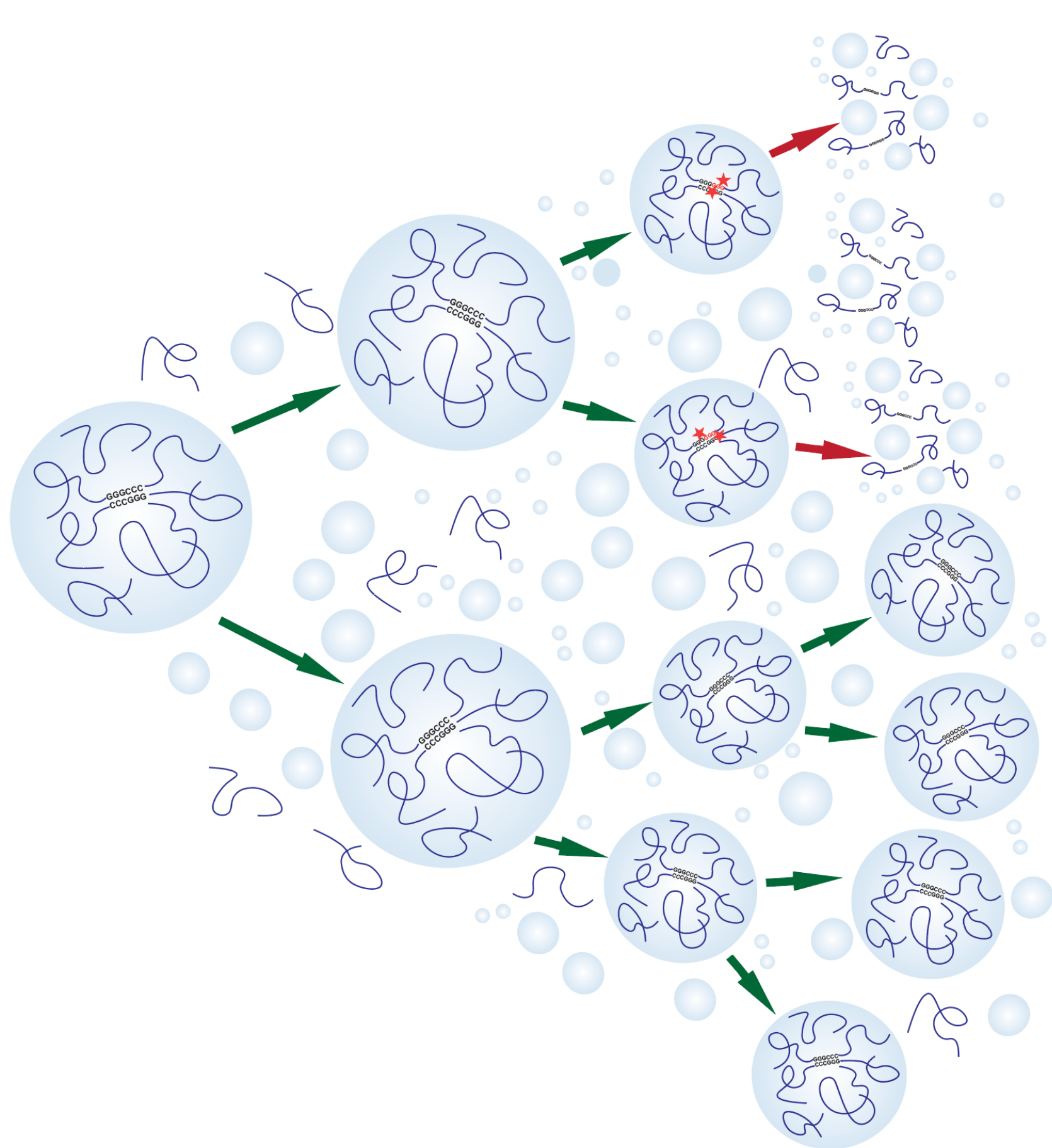

**Figure 2.** The condensate chain reaction. Exponential amplification of RNA condensates undergoing growth and division via self-templated RNA polymerization. Mutations may arise (red stars) which abolish phase separation (red arrows) in some lineages. Since these lineages no longer condense and self-replicate, natural selection for sequences that promote condensation and replication arises spontaneously.

***Explanatory power of the RNA condensate model***

One key advantage of our model is that it provides a physical basis for both compartmentalization and replication, in one unified mechanism. In other words, the formation of a system-surroundings barrier is mediated by the same polymer that produces a system capable of catalyzing its own replication.

We also suggest that the RNA-rich phase separated regions of extant cells are relics of life's origins. Potential candidates for the vestigial original condensate are the nucleolus[34,35] or the recently described dense clusters of RNA transcription in bacteria,[36] both of which are believed to be phase-separated. Interestingly, the nucleolus is a spontaneously demixed compartment, forms charge-gradients and is associated with RNA catalysis and polymerization, suggesting it may be an evolutionary relic of the model we propose. That the mitochondrial RNA polymerase POLMRT has been found to result in condensate formation[37] may suggest that the mitochondrial nucleoid is a condensate relic of the origins of such processes, also implicating phase separation in governing the transition out of the RNA world to DNA-based information storage. The latter would be consistent with the hypothesis that RNA transcription and DNA replication have a common origin, where primase and DNA polymerase descend from an ancient primase-like enzyme resemblant of the mitochondrial RNA polymerase POLRMT,[5] which carries out both functions.

In addition, if self-templated polymerization is possible using only the free energy of de-mixing (discussed further below), there is no need for an energy source such as metabolism or photosynthesis. The free energy needed for life has already been stored through an abiotic polymerization process and is accessed through the change in entropy of condensate re-mixing. Thus, the RNA condensate model could also unify metabolism with compartmentalization and replication. Remarkably, by unifying compartmentalization, energy, replication and inheritance, natural selection appears to spontaneously arise in a condensate chain reaction. Thus, the self-replicating RNA condensates can meet most of the requirements for living things.

Another key advantage of our model is that, since phase separation has been observed for polymers as short as 10 monomers[38] it may reduce the lower bound on the length of the first genes required for catalytic function. This allows for greater error thresholds, unlike the known examples of artificial RdRp ribozymes which require long sequences (>100 bases) that produce complex and stable structures. Moreover, sequence-specificity is not a strict requirement of our model. Short and low-complexity RNA sequences, if they promote the condensate chain reaction, are expected to offer a selective advantage.

The self-replicating RNA condensate of our model exhibits self-organization and emergent properties, resemblant of 'autocatalytic sets' hypothesized to model origin of life self-replicating polymers.[39] As with autocatalytic sets, the RNA polymers are marginally non-catalytic but jointly catalytic when phase separated, where each condensate catalyzes its own reproduction, and the replication of its constituent polymers.

***Hypothesized mechanism for condensate-driven polymerization.***

Here, we hypothesize that spontaneously formed condensates facilitated template-based polymerization in the RNA World, both kinetically and thermodynamically. In terms of kinetics, i.e., the condensate acting as a catalyst, we predict that the electric potential at the condensate surface functioned to reduce the activation energy of the polymerization reaction. In this reaction, polymer templates would be supplied by RNA polymers within the dense phase, while monomers and oligomers produced abiotically would be oriented[33] to the templates according to the electric field at the surface. The surface may also have functioned to stabilize reactants, by reducing degrees of freedom by positioning polymers on a two-dimensional surface,[27] rather than having them interact in three-dimensional space. This would have facilitated the exposure of the hydroxyl group of the RNA to nucleophilic attack,[40] analogous to the polymerization mechanism in extant polymerases.

During this template-based polymerization process, the bias toward which monomer is added would be physically directed by an extant polymer at the surface of the condensate. Of course, we expect the bias of a simple condensate to be weak relative to the polymerases of extant life, and inexact copies of the polymers will be made: these are replication 'errors'. However, in considering these inexact copies, we imagine that copies of the polymers with too many errors will have insufficient self-self interaction energy, and will not join the condensate. Hence the condensate has a natural 'error-correcting' mechanism that will select polymers that better resemble self.

A very weak or even non-existent bias in monomer selection during template-based polymerization could be supplemented by a natural selection for polymers that favor phase separation. Thus, even in the case where no physical template is required for the copying of the polymer and the condensate simply catalyzes template-independent polymerization, since condensate formation is sequence-dependent, polymers produced that better resemble polymers within the condensate would be more likely to be preserved over time and thereby produce more self-copies. Sequence bias within the condensate, relative to a random distribution of polymers, is therefore maintained by selection for sequences more likely to phase separate, rather than a bias due to a physical replication template.

The thermodynamics of polymerization (i.e., the change in free energy required for polymerization) would be assisted by the fact that at the surface of the RNA condensate, where we hypothesize polymerization is occurring, the reaction quotient is reduced by the increased local concentration of reactants, predicted to congregate at the surface. By Le Chatelier's principle,[41] this is expected to shift the reaction to products, i.e., producing more RNA polymers. Condensates can also act to 'exclude' products, decreasing their effective concentration, which favors the formation of more products.[27] Moreover, the crowded environment within the condensate, where constituent polymers have less degrees of conformational freedom, may lower the entropic cost of producing new polymers that join the condensate. Another source for the free energy of polymerization may have been catabolic reactions that also occurred within the condensate, such that the breakdown of higher energy molecules within the condensate may have been coupled with the formation of new polymers in a sequence-biased manner.

### *Theoretical framework for a replicative RNA condensate*

We aim to use standard concepts in thermodynamics, polymer physics, and information theory to develop a condensate-centric theoretical framework for the origin of life. We therefore first review basic polymer physics theory of condensation. Formation of a biomolecular condensate is a spontaneous process when mixing is thermodynamically unfavorable, which occurs if the change in free energy of mixing ($\Delta G_{mix}$), a function of enthalpy ($\Delta H_{mix}$), temperature ($T$) and entropy ($\Delta S_{mix}$), is greater than zero (Equation 2).

$$\Delta G_{mix} = \Delta H_{mix} - T\Delta S_{mix}$$

(Equation 3)

The Flory-Huggins theory[42,43] approximates the free energy of phase separation for a binary mixture of polymers and solvent. The polymer-solvent system is modelled as a three-dimensional lattice, where polymer units compete with solvent molecules for lattice sites, and sites are either occupied by an individual monomer or solvent molecule. If there are $N$ monomers per polymer, and $n = 1$ lattice sites filled per solvent molecule, with $\varphi$ and $\phi$ volume fractions for solvent and polymers ($\varphi + \phi = 1$), respectively, where $k_B$ is Boltzman's constant, $T$ is temperature, and $\chi$ is the interaction parameter, then the Gibbs free energy of mixing can be expressed as follows:

$$\Delta G_{mix} = k_B T \left[ (1-\phi)\ln(1-\phi) + \frac{\phi}{N}\ln\phi + \chi\phi(1-\phi) \right]$$

(Equation 4)

For the origins of the RNA World from prebiotic chemistry, we desire $N$ and $\phi$ to be as small as possible (Table 1). This arises because (i) larger $N$ values require lower error rates by Eigen's error threshold and (ii) larger $\phi$ values require more polymers to exist per unit of volume in the primordial soup, which requires more complex prebiotic chemistry. Notably, condensation has been observed for mixtures of RNA and proteins as short as 10 monomers.[38]

RNA polymers are likely to exist in solutions which may contain sodium, magnesium or other counterions with positive charges. For example, Voorn-Oberbeek theory[44] adds free energy terms for long-range Columbic interactions, from which a closed-form expression for $\chi$ may also be obtained (e.g., Equation 19 from Veis (2011).). We also note that electrostatic interactions may also drive phase-separation, if positive counterions are wedged between negative polymers, producing an effective attractive force between polyelectrolytes. Indeed, $Mg^{2+}$ has been shown to drive the phase separation of RNA.[46] Generally, the phase separation of RNA as a polymer is unique as it is governed by base-pairing interactions[46,47] as well as the nucleotide composition of interacting polymers[48] and secondary structure.[49] In the current cellular context, repetitive and GC-rich sequences are linked to greater phase separation.[49]

In the case of an RNA sequence made from an alphabet of four bases {A, U, C, G}, a simple approximation of $\chi$ is a function of the energies of all pairs of monomer-monomer interactions within a polymer of length $N$ (Equation 5).[50] The energy, $\varepsilon_{r_\alpha, r_\beta}$ is given by a 4x4 matrix representing the energy of all possible base-base interactions, where $r_\alpha$ and $r_\beta$ denote the bases observed at the $\alpha^{th}$ and $\beta^{th}$ position in the polymer.

$$\chi \propto \frac{1}{N^2} \sum_{\alpha=1}^{N} \sum_{\beta=1}^{N} \varepsilon_{r_\alpha, r_\beta}$$

(Equation 5)

In this simplification, the values of $\chi$ in Equation 5 are not impacted by the order of bases, rather, they depend only on the base composition.[48] By averaging across pairwise interactions, the model does not consider that at any moment in time, only a subset of possible interactions can be formed by hydrogen bonding between bases on interacting polymers. Importantly, this model of RNA-RNA interactions is position-independent and discards local effects of secondary structure that may arise from non-Watson-Crick base pairing and base stacking, which may also depend on salt concentration and temperature. These features are better captured by the Turner rules,[51] and are employed in computational models of RNA structure.[52] To our knowledge, there is no widely accepted theory of nucleic acid condensate that considers base order (see directions for future research below), but the RNA base-paring and stacking interactions can be approximated by a large $\chi$ parameter, because in a spontaneously condensing RNA polymer, they are likely to represent strong self-self interactions.[53]

Given the free energy of mixing of a condensate, we next consider whether this free energy might be sufficient for life. It is widely appreciated that life must be a non-equilibrium, dissipative process,[16] but to our knowledge, there is no specific understanding of what the minimal dissipative steps are. Our RNA condensate model is simple enough that we can formalize the minimum free energy cost of replication. As the simplest case, we consider the origin of life as the process by which a population of uniformly distributed RNA polymers produced at random by the abiotic environment evolves into a non-uniform population of condensate compartmentalized RNAs. In this minimal situation, we imagine the condensate is catalyzing only swapping of monomers from one polymer to another, so the enthalpic and entropic costs of polymerization are borne by the abiotic polymerization processes. Thus, the only free energy cost is the reduction in entropy due to the reduction in diversity when abiotic background RNA sequences are converted to condensate-specific RNA sequences. We ask whether free energy of mixing due to condensate formation is sufficient to maintain a sequence-biased, and therefore non-uniform, information-rich, population of RNAs.

For a formal description of how the free energy of condensate formation relates to the entropy cost of maintaining a population of information-rich polymers in the origin of life, let us consider a population of RNA polymers of length $N$ produced by the abiotic environment. Assuming the sequences are unbiased, then the "background" distribution of RNA polymers, $X$ is given by $X \sim U(0, 4^N)$, which we denote as $Q(X) = 1/4^N$. Following Eigen, we imagine that there exists a single 'master' polymer sequence capable of condensation and self-replication. Since replication is likely to be error prone, the probability that any replicated polymer will differ by $d$ bases (the Hamming distance) from the master sequence with mutation rate $r$ is $d \sim Binom(r, N)$. The probability of each polymer sequence in terms of the distribution of differences form the master, by the law of total probability, is $P(X) = \sum_{d=0}^{N} P(X|d)P(d)$. Because number of RNA sequences $d$ away from the master is $\binom{N}{d}3^d$, we use $P(X|d) = 1/\binom{N}{d}3^d$ as the probability of a sequence, given $d$ differences from the master. Since we assume there are no enthalpic or conformational entropy costs of because the RNA polymers are all of equal length, the minimum free energy needed for the origin of life is:

$$\Delta G_{min} = -T\Delta S = T(S[Q(X)] - S[P(X)])$$

(Equation 6)

Where $S[p(x)] = -k_b \sum_x p(x) \log p(x)$ is the Gibbs entropy. Thus, the difference $S[Q(X)] - S[P(X)]$ can be interpreted as a loss of entropy from having a sequence-biased population of polymers, with respect to the abiotic background. Since $P(X)$ is a function of the mutation rate, $r$, we are interested in the region where the error rate is below Eigen's threshold $r < 1/N$, because mutation rates higher than $1/N$ lead to error catastrophes. In other words, the value $\Delta G_{min}$ that we obtain at $r = 1/N$ is the minimum free energy the origin of life requires to proceed.

We can now compare the free energy required to exceed Eigen's threshold ($\Delta G_{min}$), computed $P(x)$ and $Q(x)$ as above, to the free energy of condensate formation under a Flory-Huggins model. In Figure 3a, we plot $\Delta G_{mix}/k_B T$ as a function of volume fraction $\phi$ and interaction parameter $\chi$ using Equation 4, and in Figure 3b we show $\Delta G_{min}/k_B T$ as a function of polymer length $N$ using Equation 6 for the case of no errors $r = 0$ and Eigen's threshold $r = 1/N$. Intriguingly, the free energy values obtained by plotting the Flory-Huggins model for different polymer lengths and interaction parameter values in Figure 3a are, for high values of $\chi$, within similar orders of magnitude as the values obtained in Figure 3b. Whether these values of $\chi$ are reasonable within real-world scenarios of RNA condensation remains to be determined.

For a given $\chi, \phi$, the point where $\Delta G_{mix} = \Delta G_{min}$at Eigen's threshold $(r = 1/N)$ identifies the maximum length of a self-replicating master sequence that can be sustained using the free energy available in the condensate. To get a sense of the maximum polymer length, we plot $\Delta G_{mix}$ for $\chi \in \{5,20,50\}$ for $\phi = 0.3$ in Figure 3b. The intersection of curves predicts that polymers of maximum 2, 5, and 10 monomers could form for $\chi \in \{5,20,50\}$ during the origin of life, if the free energy of condensate formation is used.

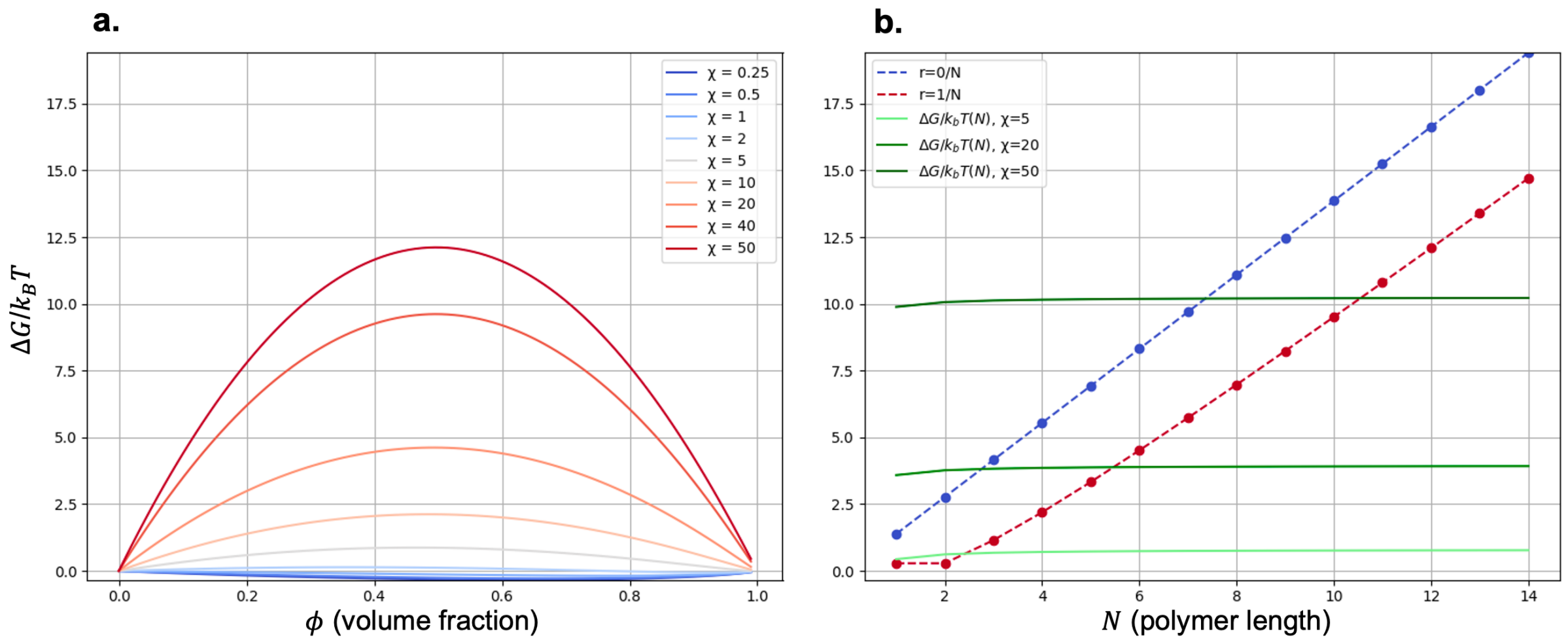

**Figure 3. a)** Free energy of mixing $\Delta G_{mix}/k_B T$ at varying polymer fractions, under the Flory-Huggins model, for different $\chi$ values with $N = 10$ and **b)** Minimum free energy of the origin of life $\Delta G_{min}/k_B T$ sequence length $N$ for mutation rate 0 or $1/N$. The 'zone of life' is the area between the two free energy curves. Values from $\Delta G_{mix}/k_B T$ are plotted in panel **b.** as a function of $N$ such that $\chi \in \{5,20,50\}$, and $\phi = 0.3$, to illustrate polymer lengths that are admissible under Eigen's error threshold for different values of $\chi$.

In Table 1, we summarize the key biophysical parameters of our RNA condensate model discussed throughout the paper, and how they address key aspects of the origin of life. Our model provides a useful framework in which to formalize the origin of life in terms of parameters that relate to our RNA condensate model: compartmentalization is addressed by $\chi$, the shortest polymer length and its concentration is addressed by $N$ and $\phi$, respectively, and Eigen's error rate is related to $r$ and the minimum free energy required for a population of RNA polymers to replicate above Eigen's error rate is given by $\Delta G_{min}$.

**Table 1. Five key biophysical parameters that address core aspects of the origin of life.**

| Model parameter | Description | Interpretation | Relationship with origin of life | Key equations |
|---|---|---|---|---|
| $\chi$ | Interaction parameter | Phase separation occurs when value exceeds minimum threshold | Addresses compartmentalization; no membrane needed | 2,3,5 |
| $N$ | Number of monomers per polymer | Shorter polymers easier to form from prebiotic chemistry | Places lower bound on length of first genes | 3 |
| $\phi$ | Volume fraction occupied by polymers in solution | Smaller volume fractions easier to form from prebiotic chemistry | Place lower bound on volume occupied of first genes | 3 |
| $r$ | Error rate of self-templated polymerization | RNA condensate self-replication error rate | No need for highly ordered enzyme structure; reduces error threshold | 1 |
| $\Delta G_{min}$ | Minimal free energy to pass Eigen's error threshold | More information requires more free energy | Minimum work required from condensate formation for the origin of life | 6 |

### *Plausibility and possible technical considerations*

Water is often considered a good solvent for single stranded nucleic acid sequences, and therefore they are not usually expected to form condensates. However, RNA-sequence encoded RNA condensates have recently been achieved experimentally: condensation of so-called 'RNA-stars'[54] that can be controlled by canonical base-pairing of only six bases. In this work, the authors show that these simple base-pairing interactions are sufficient to achieve condensate specificity, such that condensates encoded by differing RNA sequences do not mix.

An important technical issue that arises during condensate-catalyzed RNA-replication is whether the canonical base-pairing required for template-mediated polymerization would lead to the formation of stable double-stranded RNA. If so, the multivalent interactions that form the condensate would need to be strong enough to overcome this tendency. The formation of double-stranded intermediates is known to occur during both DNA-templated RNA polymerization (transcription) and RNA-templated DNA transcription (reverse transcription). On the other hand, when phase separated, RNA molecules have also been observed to become immobile, constituting 'dynamical arrest'.[48,55] Similarly, if $\chi$ is too large, the condensate may undergo continued growth preventing division, which would also reduce the relative surface area and perhaps slow down the catalysis. Thus, there may be some critical range of $\chi$ where condensation is optimal, while enabling condensates to break apart rapidly enough to leave the interface available for driving catalysis. It is also reasonable to hypothesize that temperature cycles and/or physical forces can work to shear apart condensates that grow too large, resembling cycles of model protocell growth and division.[19]

### *Directions for origin of life research on RNA condensates*

Key next steps for the physical plausibility of the RNA-condensate origin of life model would be (i) to identify short and low-complexity RNA sequences that can form condensates, and (ii) whether the free energy available in the formed RNA condensates would be sufficient to overcome the activation energies needed catalyze polymerization of RNA sequences of that length. These questions are well within the reach of current experimental and theoretical capability. Key to these steps would be an understanding of polymer-polymer interactions that considers the order of bases in RNA sequences. It is perhaps most crucial to explore whether newly produced RNAs are of sufficient character to elicit phase separation, and whether mixing-demixing cycles can produce RNA condensates that transmit minimal genetic information (RNA composition and/or sequence) to 'offspring' condensates.

The simplicity of the RNA condensate model allows us to consider formally the origins of molecularly encoded biological information.[5] The reduction of RNA sequence diversity due to replication of polymers more similar to the template (the polymers within the condensate) reduces the entropy of the polymers and therefore costs free energy. This free energy cost of replication can be interpreted as the free energy required for the transmission of sequence information from the parent polymer to its offspring. By extrapolating from Landauer's principle[56] we therefore have a lower bound on the minimum amount of free energy required for encoding some number of bits $I$ into the population of RNA sequences within the condensate: $\Delta G_{min} = I k_b T \ln 2$, so that every bit of information (here, sequence information) requires at least $k_b T \ln 2$ joules of energy. Thus, we can convert the minimum free energy into a maximum amount of information that can be transmitted using only the free energy of demixing. We anticipate that this simple physical connection of replication with information theory can be used to gain further insight into the origin of life. Of course, we imagine that once the condensate chain reaction begins, the condensates will evolve to exploit other sources of free energy in their environment, for instance, hydrothermal vents[57] and ultraviolet radiation.[58]

To better frame these questions, further research into elucidating the mechanisms of condensate-based catalysis would be very helpful. Altogether, addressing these issues and providing support for the model we present is expected to resolve key issues in origin of life research with implications for understanding extant life.

***Acknowledgements***

We thank Drs. Jonas Wessén, Tanmoy Pal, Wang Tat Yau, and Hue Sun Chan for their expert feedback on our model and for their critical reading of the manuscript. We thank Ami Sangster, Alex Tianhao Huang and Dr. Tony Harris for comments on the manuscript. We acknowledge Drs. Rohit Pappu, Alex Holehouse, Gabrielle Abraham, Jane Dyson and Julie Forman-Kay for inspiring discussions. We thank Dr. Rohit Pappu and an anonymous reviewer for helpful comments and suggestions.

***References***

1. Gilbert, W. Origin of life: The RNA world. *Nature* **319**, 618 (1986).

2. Kruger, K. *et al.* Self-splicing RNA: autoexcision and autocyclization of the ribosomal RNA intervening sequence of Tetrahymena. *Cell* **31**, 147–157 (1982).

3. Guerrier-Takada, C., Gardiner, K., Marsh, T., Pace, N. & Altman, S. The RNA moiety of ribonuclease P is the catalytic subunit of the enzyme. *Cell* **35**, 849–857 (1983).

4. Nissen, P., Hansen, J., Ban, N., Moore, P. B. & Steitz, T. A. The structural basis of ribosome activity in peptide bond synthesis. *Science* **289**, 920–930 (2000).

5. Fine, J. & Pearlman, R. On the origin of life: an RNA-focused synthesis and narrative. *RNA* **29**, rna.079598.123 (2023).

6. Cech, T. R. The RNA worlds in context. *Cold Spring Harb Perspect Biol* **4**, a006742 (2012).

7. Strobel, S. A. & Cochrane, J. C. RNA catalysis: ribozymes, ribosomes, and riboswitches. *Curr Opin Chem Biol* **11**, 636–643 (2007).

8. Bernier, C. R., Petrov, A. S., Kovacs, N. A., Penev, P. I. & Williams, L. D. Translation: The Universal Structural Core of Life. *Mol Biol Evol* **35**, 2065–2076 (2018).

9. Petrov, A. S. *et al.* Evolution of the ribosome at atomic resolution. *Proceedings of the National Academy of Sciences* **111**, 10251–10256 (2014).

10. Crick, F. H. The origin of the genetic code. *J Mol Biol* **38**, 367–379 (1968).

11. Johnston, W. K., Unrau, P. J., Lawrence, M. S., Glasner, M. E. & Bartel, D. P. RNA-catalyzed RNA polymerization: accurate and general RNA-templated primer extension. *Science* **292**, 1319–1325 (2001).

12. Zaher, H. S. & Unrau, P. J. Selection of an improved RNA polymerase ribozyme with superior extension and fidelity. *RNA* **13**, 1017–1026 (2007).

13. Horning, D. P. & Joyce, G. F. Amplification of RNA by an RNA polymerase ribozyme. *Proceedings of the National Academy of Sciences* **113**, 9786 LP-- 9791 (2016).

14. Cojocaru, R. & Unrau, P. J. Processive RNA polymerization and promoter recognition in an RNA World. *Science (1979)* **371**, 1225–1232 (2021).

15. Tagami, S., Attwater, J. & Holliger, P. Simple peptides derived from the ribosomal core potentiate RNA polymerase ribozyme function. *Nat Chem* **9**, 325–332 (2017).

16. Eigen, M. Selforganization of matter and the evolution of biological macromolecules. *Naturwissenschaften* **58**, 465–523 (1971).

17. Papastavrou, N., Horning, D. P. & Joyce, G. F. RNA-catalyzed evolution of catalytic RNA. *Proceedings of the National Academy of Sciences* **121**, e2321592121 (2024).

18. Kun, Á., Santos, M. & Szathmáry, E. Real ribozymes suggest a relaxed error threshold. *Nat Genet* **37**, 1008–1011 (2005).

19. Zhu, T. F. & Szostak, J. W. Coupled Growth and Division of Model Protocell Membranes. *J Am Chem Soc* **131**, 5705–5713 (2009).

20. Zhou, L., O'Flaherty, D. & Szostak, J. Template-Directed Copying of RNA by Non-enzymatic Ligation. *Angewandte Chemie International Edition* **59**, (2020).

21. Jin, L., Kamat, N. P., Jena, S. & Szostak, J. W. Fatty Acid/Phospholipid Blended Membranes: A Potential Intermediate State in Protocellular Evolution. *Small* **14**, e1704077 (2018).

22. Garcia-Jove Navarro, M. *et al.* RNA is a critical element for the sizing and the composition of phase-separated RNA–protein condensates. *Nat Commun* **10**, 3230 (2019).

23. Alberti, S. Phase separation in biology. *Current Biology* **27**, R1097--R1102 (2017).

24. Poudyal, R. R., Pir Cakmak, F., Keating, C. D. & Bevilacqua, P. C. Physical Principles and Extant Biology Reveal Roles for RNA-Containing Membraneless Compartments in Origins of Life Chemistry. *Biochemistry* **57**, 2509–2519 (2018).

25. Kolb, V. M. Oparin's coacervates as an important milestone in chemical evolution. in *Proc.SPIE* vol. 9606 (2015).

26. Stroberg, W. & Schnell, S. Do Cellular Condensates Accelerate Biochemical Reactions? Lessons from Microdroplet Chemistry. *Biophys J* **115**, 3–8 (2018).

27. Smokers, I. B. A., Visser, B. S., Slootbeek, A. D., Huck, W. T. S. & Spruijt, E. How Droplets Can Accelerate Reactions─Coacervate Protocells as Catalytic Microcompartments. *Acc Chem Res* **57**, 1885–1895 (2024).

28. Alston, J. J. & Soranno, A. Condensation Goes Viral: A Polymer Physics Perspective. *J Mol Biol* **435**, 167988 (2023).

29. Dai, Y. *et al.* Interface of biomolecular condensates modulates redox reactions. *Chem* **9**, 1594–1609 (2023).

30. Guo, X. *et al.* Biomolecular condensates can function as inherent catalysts. *bioRxiv* 2024.07.06.602359 (2024) doi:10.1101/2024.07.06.602359.

31. King, M. R. *et al.* Macromolecular condensation organizes nucleolar sub-phases to set up a pH gradient. *Cell* **187**, 1889-1906.e24 (2024).

32. Posey, A. E. *et al.* Biomolecular Condensates are Characterized by Interphase Electric Potentials. *J Am Chem Soc* (2024) doi:10.1021/JACS.4C08946.

33. Dai, Y., Wang, Z.-G. & Zare, R. N. Unlocking the electrochemical functions of biomolecular condensates. *Nature Chemical Biology 2024 20:11* **20**, 1420–1433 (2024).

34. Lafontaine, D. L. J., Riback, J. A., Bascetin, R. & Brangwynne, C. P. The nucleolus as a multiphase liquid condensate. *Nature Reviews Molecular Cell Biology 2020 22:3* **22**, 165–182 (2020).

35. Jaberi-Lashkari, N., Lee, B., Aryan, F. & Calo, E. An evolutionarily nascent architecture underlying the formation and emergence of biomolecular condensates. *Cell Rep* **42**, 112955 (2023).

36. Ladouceur, A. M. *et al.* Clusters of bacterial RNA polymerase are biomolecular condensates that assemble through liquid-liquid phase separation. *Proc Natl Acad Sci U S A* **117**, 18540–18549 (2020).

37. Feric, M. *et al.* Mesoscale structure–function relationships in mitochondrial transcriptional condensates. *Proceedings of the National Academy of Sciences* **119**, e2207303119 (2022).

38. Valdes-Garcia, G. *et al.* The effect of polymer length in liquid-liquid phase separation. *Cell Rep Phys Sci* **4**, (2023).

39. Kauffman, S. The Origins of Order: Self-Organization and Selection in Evolution. *emergence.org* **15**, (1992).

40. Aranda, J., Wieczór, M., Terrazas, M., Brun-Heath, I. & Orozco, M. Mechanism of reaction of RNA-dependent RNA polymerase from SARS-CoV-2. *Chem Catalysis* **2**, 1084 (2022).

41. Hillert, M. Le Chatelier's principle-restated and illustrated with phase diagrams. *Journal of Phase Equilibria* **16**, 403–410 (1995).

42. Flory, P. J., Flory & J., P. Thermodynamics of High Polymer Solutions. *JChPh* **10**, 51–61 (1942).

43. Huggins, M. L. Solutions of Long Chain Compounds. *JChPh* **9**, 440–440 (1941).

44. OVERBEEK, J. T. & VOORN, M. J. Phase separation in polyelectrolyte solutions. Theory of complex coacervation. *J Cell Comp Physiol* **49**, 7–26 (1957).

45. Veis, A. A REVIEW OF THE EARLY DEVELOPMENT OF THE THERMODYNAMICS OF THE COMPLEX COACERVATION PHASE SEPARATION. *Adv Colloid Interface Sci* **167**, 2 (2011).

46. Ma, Y. *et al.* Nucleobase Clustering Contributes to the Formation and Hollowing of Repeat-Expansion RNA Condensate. *J Am Chem Soc* **144**, 4716–4720 (2022).

47. Abraham, G. R., Chaderjian, A. S., N Nguyen, A. B., Wilken, S. & Saleh, O. A. Nucleic acid liquids. *Rep Prog Phys* **87**, (2024).

48. Wadsworth, G. M. *et al.* RNAs undergo phase transitions with lower critical solution temperatures. *Nat Chem* **15**, 1693–1704 (2023).

49. Wang, S. & Xu, Y. RNA structure promotes liquid-to-solid phase transition of short RNAs in neuronal dysfunction. *Communications Biology 2024 7:1* **7**, 1–14 (2024).

50. Wessén, J., Das, S., Pal, T. & Chan, H. S. Analytical Formulation and Field-Theoretic Simulation of Sequence-Specific Phase Separation of Protein-Like Heteropolymers with Short- and Long-Spatial-Range Interactions. *Journal of Physical Chemistry B* **126**, 9222–9245 (2022).

51. Turner, D. H. Thermodynamics of base pairing. *Curr Opin Struct Biol* **6**, 299–304 (1996).

52. Zuber, J., Schroeder, S. J., Sun, H., Turner, D. H. & Mathews, D. H. Nearest neighbor rules for RNA helix folding thermodynamics: improved end effects. *Nucleic Acids Res* **50**, 5251–5262 (2022).

53. Abraham, G. R., Chaderjian, A. S., N Nguyen, A. B., Wilken, S. & Saleh, O. A. Nucleic acid liquids. *Rep Prog Phys* **87**, (2024).

54. Fabrini, G. *et al.* Co-transcriptional production of programmable RNA condensates and synthetic organelles. *Nature Nanotechnology 2024 19:11* **19**, 1665–1673 (2024).

55. Lin, A. Z. *et al.* Dynamical control enables the formation of demixed biomolecular condensates. *Nature Communications 2023 14:1* **14**, 1–17 (2023).

56. Landauer, R. Irreversibility and Heat Generation in the Computing Process. *IBM J Res Dev* **5**, 183–191 (2010).

57. Martin, W., Baross, J., Kelley, D. & Russell, M. J. Hydrothermal vents and the origin of life. *Nature Reviews Microbiology 2008 6:11* **6**, 805–814 (2008).

58. Saha, R. & Chen, I. A. Effect of UV Radiation on Fluorescent RNA Aptamers' Functional and Templating Ability. *Chembiochem* **20**, 2609–2617 (2019).